\documentstyle[epsf]{cupconf}

\ifoldfss
\else
  \ifnfssone
    \newmathalphabet{\mathit}
      \addtoversion{normal}{\mathit}{cmr}{m}{it}
      \addtoversion{bold}{\mathit}{cmr}{bx}{it}
    \newmathalphabet{\mathcal}
      \addtoversion{normal}{\mathcal}{cmsy}{m}{n}
    \else
    \ifnfsstwo
    \fi
  \fi
\fi

\newcommand{\la}{\mathrel{\mathchoice{\vcenter{\offinterlineskip\halign{\hfil
$\displaystyle##$\hfil\cr<\cr\sim\cr}}}
{\vcenter{\offinterlineskip\halign{\hfil$\textstyle##$\hfil\cr<\cr\sim\cr}}}
{\vcenter{\offinterlineskip\halign{\hfil$\scriptstyle##$\hfil\cr<\cr\sim\cr}}}
{\vcenter{\offinterlineskip\halign{\hfil$\scriptscriptstyle##$\hfil\cr<\cr\sim\cr}}}}}

\def\HI{\mbox{H{\scriptsize I}}}
\def\ie{{i.e.}}
\def\eg{{e.g.}}
\def\etal{{et al.}}
\def\Msol{\mbox{M$_\odot$}}
\newcommand{\Lya}{Ly$\alpha$\ }

\def\aaa{A\&A}
\def\apj{ApJ}

\def\aj{AJ}
\def\mnras{MNRAS}
\def\nat{Nature}

\title{Model Predictions for Clustering and Morphologies at HDF depths}

\author[M.~Steinmetz]%
{M\ls A\ls T\ls T\ls H\ls I\ls A\ls S\ns S\ls T\ls E\ls I\ls N\ls M\ls E\ls T\ls Z}

\affiliation{Steward Observatory, University of Arizona, Tucson, AZ 85721, USA}

\begin{document}
\ifnfssone
\else
  \ifnfsstwo
  \else
    \ifoldfss
      \let\mathcal\cal
      \let\mathrm\rm
      \let\mathsf\sf
    \fi
  \fi
\fi

\maketitle

\begin{abstract}
The current status of numerical simulations of the formation of galaxies
is reviewed. Success and failure of modeling galaxies
at low and high redshift is demonstrated using a variety of examples, such as
the Tully-Fisher relation, the appearance of high-redshift galaxies and the
kinematics of damped \Lya systems. The relationship between the
clustering properties of high-$z$ galaxies and the present 
generation of galaxies is emphasized.

\end{abstract}

\firstsection % if your document starts with a section,
\section{Introduction}

Hierarchical clustering is at present the most successful paradigm of structure
formation in the universe. In this scenario -- Cold Dark Matter (CDM) and its variants are
perhaps the best known examples -- structure grows as systems of
progressively larger mass merge and collapse to form newly
virialized systems. A large variety of models which are based on the
hierarchical clustering hypothesis have been extensively studied using N-body
simulations as well as analytical approximations.  

Within the last few years, more and more work focussed on smaller structures,
and tried to embed galaxy formation into the hierarchical picture. One
successful approach has been semi-analytical or phenomenological models
(Kauffmann, White \& Guiderdoni, 1993; Cole \etal, 1994). These models use the
extended Press-Schechter theory to predict abundances and merger rates of
halos as a function of mass and redshift. Physically motivated recipes
are used to model how gas cools, how it settles at the center of dark
matter halos and how it is transformed into stars. This phenomenological ansatz
provides at comparably low computational cost a very efficient method to
predict the formation and evolution of the galaxy population. However, it has only
little power in predicting the clustering  of galaxies (see, however, Kauffmann,
Nusser \& Steinmetz, 1997) and it makes no prediction 
on the detailed formation history of individual galaxies. 

Detailed studies of the formation of individual galaxies necessarily
require large numerical simulations which include not only gas dynamics, shocks
and radiative cooling, but which also incorporate some description of star
formation. This article addresses some successes and failures of such simulations.
It will focus on issues which seem to be generic to the hypothesis of
hierarchical clustering and which depend only little on the details on the
underlying cosmogony. Numerical details are discussed in the appendix.

\begin{figure}[ht]
{\hskip0.01\hsize\epsfxsize=0.48\hsize\epsffile{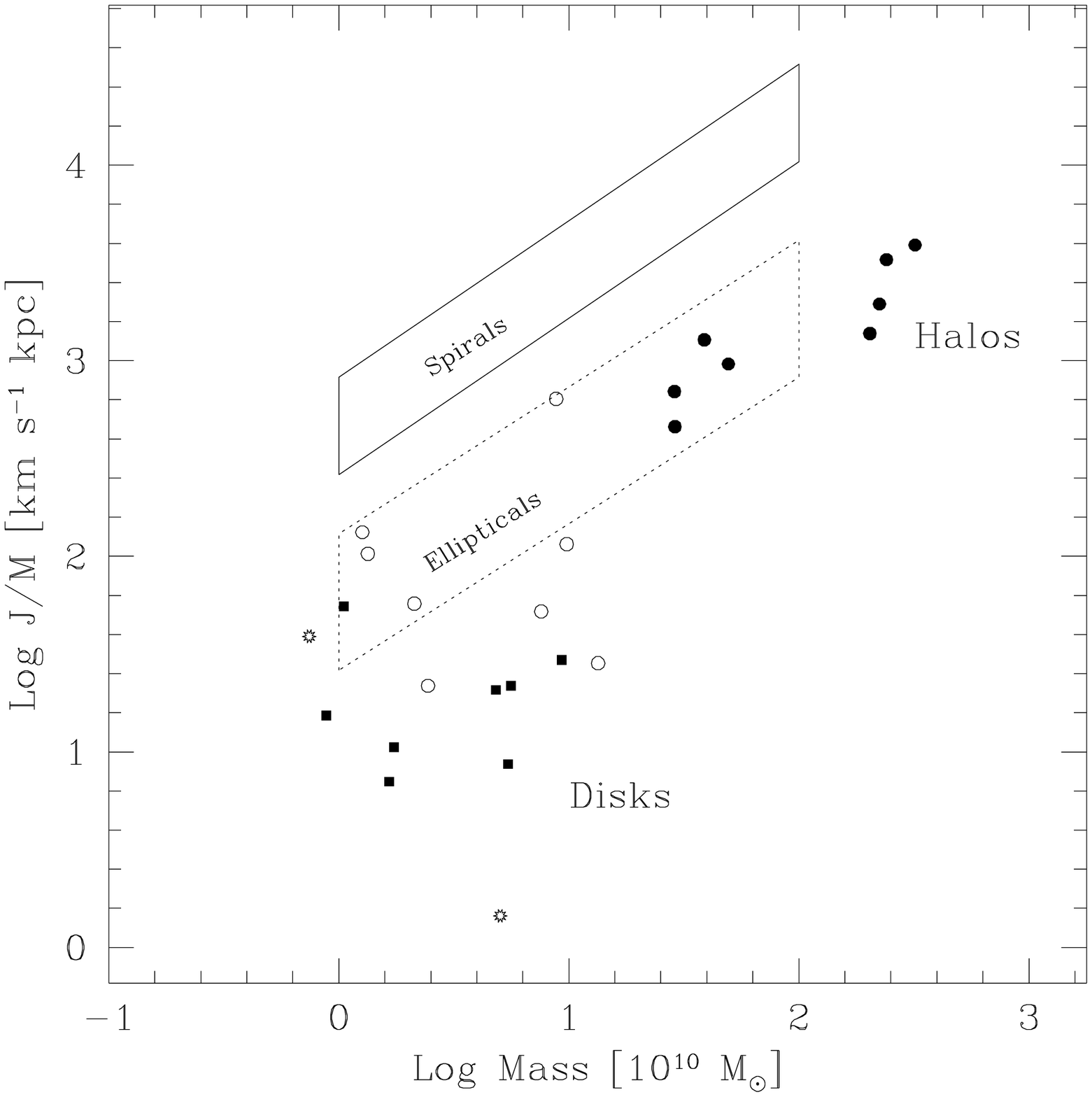}\hskip0.02\hsize\epsfxsize=0.48\hsize\epsffile{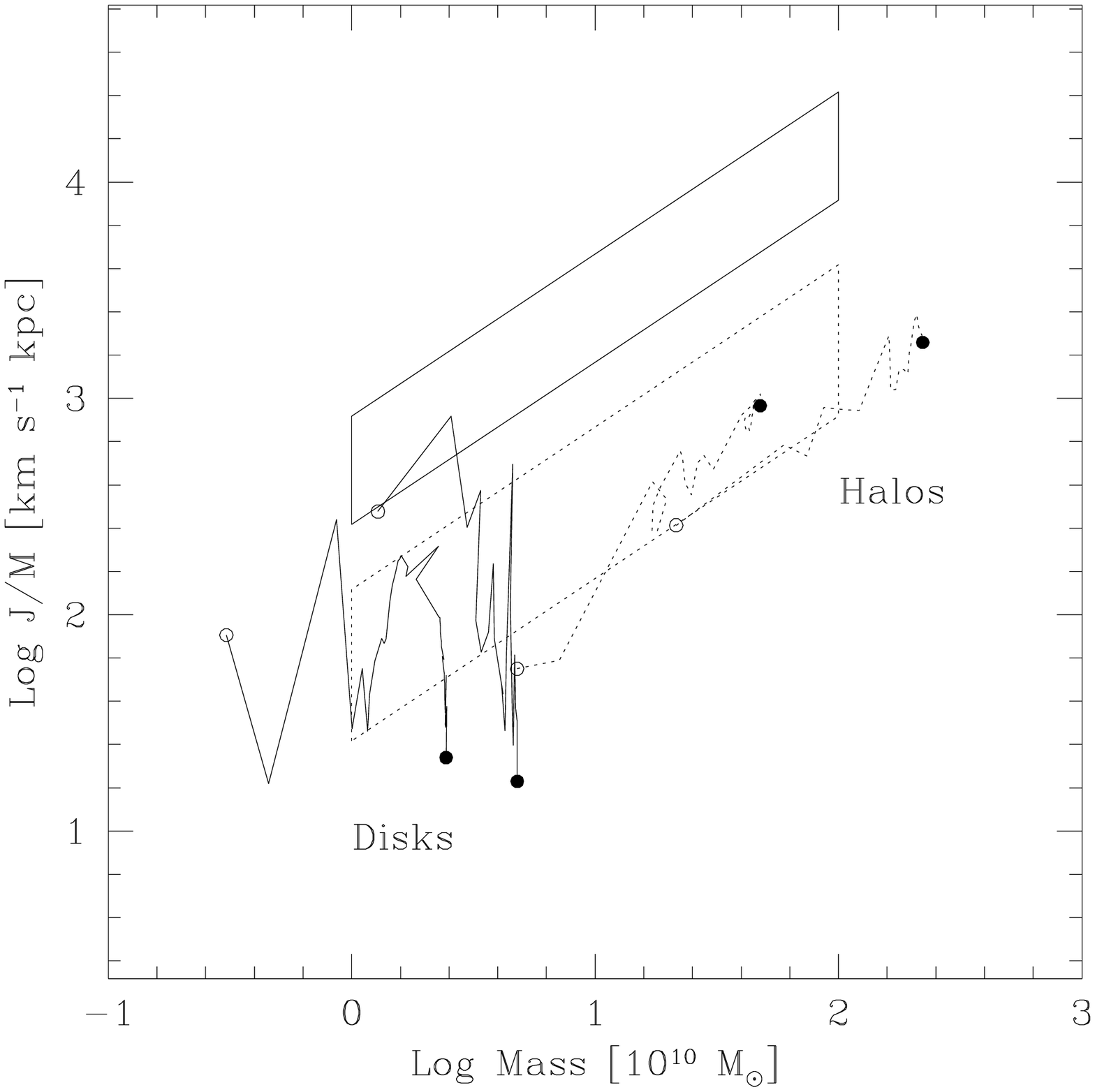}\hfill\break}
\caption{\label{ns97}Left: The specific angular momentum of dark halos and
gaseous disks, as a function of mass. The boxes enclose the region occupied by
spiral and elliptical galaxies, as given by Fall (1983). Open circles correspond
to runs without UV background; solid squares correspond to a soft UV background
($J_{-21}=1$, $\alpha=5$ (equation 2.3)), starred symbols correspond to an
extremely energetic background ($J_{-21}=10$, $\alpha=1$). Note that the halos'
$J/M$ scale approximately as $M^{2/3}$, as expected if all systems had the same
value of the rotation parameter $\lambda$ (see text for a definition). Gaseous
disks have much lower angular momenta than observed spirals, a consequence of
the role of mergers during the assembly of the disks. Note that the inclusion of
a UV radiation field seems to aggravate this problem.  Right: Evolution of the
dark halo and central gaseous disk in the $J/M$ versus $M$ plane, from $z=5$
(open circles) to $z=0$ (solid circles).  
The mass of the system grows steadily by mergers,
which are accompanied by an increase in the spin of the halo and a decrease in
the spin of the central disk. The latter results from angular momentum being
transferred from the gas to the halo during mergers.}

\end{figure}

\section{Prologue: The Overcooling Problem}

Already the very first quantitative investigations of galaxy formation in
hierarchically clustering universes (\eg, White \& Rees 1978) exhibited a generic problem of
this class of models, nowadays usually referred to as the
{\sl overcooling problem}. Since cooling times scale inversely with density, the
dissipative collapse of gas must have been more efficient at high
redshift because the dark matter halos present at that time (and the
universe as a whole) were denser. Cooling is expected to be so
efficient at early times that almost all gas within a dark matter halo is able
to cool to temperatures of about $10^4$\,K. Consequently, galaxies tend to be
too massive. A related problem is that hierarchical models predict an
overabundance of low mass halos with circular velocities below $100\,$km/s
(Cole \etal, 1994).

This can be quantified by the following analytical argument (see, \eg, White
1994). At a given redshift  $z$, the mass of a dark matter halo of circular velocity $v_c$  can be written as
\begin{equation}
M_{\rm vir} = 2.6\times 10^{12}\,\Msol\,\frac{3\zeta}{2}\,h^{-1}\,\left(1+z\right)^{-1.5}\left(\frac{v_c}{220\,\mbox{km/s}}\right)^3\,, 
\end{equation}
the factor $\zeta$ representing the age of the universe in units of $H_0^{-1}$,
\ie, for a $\Omega=1$ universe $\zeta=2/3$. Following White \etal\ (1993), the
baryon fraction ($M_{\rm bary}/M_{\rm tot}$) within the virial radius should be very
similar to the cosmological value $\Omega_{\rm b}/\Omega_0$, with the
nucleosynthesis value $\Omega_{\rm b} = 0.0125\,h^{-2}$. The
baryonic mass enclosed in a halo of circular velocity $v_c$ is thus given by

\begin{equation}
M_{\rm bary}  = \frac{\Omega_{\rm bary}}{\Omega_0} M_{\rm vir}
 =  3.3 \times 10^{10}\,\Msol\,\frac{3\zeta}{2\Omega_0}\,h^{-3}\,\left(1+z\right)^{-1.5}\,\left(\frac{v_{\rm vir}}{220
\,\mbox{km/s}}\right)^3\, .
\end{equation}
Let us apply this to the case of the Galaxy.
By assuming that the circular velocity of the dark matter halo is similar 
to the actual rotation velocity of the Galactic disk ($M_{\rm disk} \approx
6\times10^{10}\,\Msol$),the following statements can be made:
\begin{itemize}
\item In the case of $\Omega_0=1$ and $h=0.5$, the total baryonic mass amounts
$2.6\times 10^{11}\,\Msol$, \ie, 4 times the mass of the galactic disk. For
$\Omega=0.2,\Lambda= 0.8$ the situation is even more extreme, the total baryonic mass is
$2.1~10^{12}\,\Msol$, 
\ie, only 3\% of the baryonic mass has cooled and settled into the disk. 
\item The situation is a bit less extreme if a higher $h$ and/or a
lower $\Omega_{\rm b}$ is assumed, as, \eg, favored by recent Deuterium measurements by Songaila
\etal\ (1994). Vice versa, even less gas has cooled if a higher 
$\Omega_{\rm b}$ is assumed, as, \eg, favored by the Deuterium measurements of
Tytler, Fan \& Burles (1996). 
\item Since  the cooling radius of a Milky Way
sized system includes much more mass than the mass of the disk, a very efficient
heating mechanism has to be postulated to prevent a large fraction of the
baryonic mass from cooling.
\item It is of some interest to note that for $\Omega=1$ and $h=0.8$ there
exist hardly enough baryons to account for the mass observed in the galactic disk.
\end{itemize}

This simple model nicely reproduces the main feature seen in numerical
simulation of galaxy formation (see, \eg, Navarro \& Steinmetz 1997), namely
that virtually all baryons within a dark matter halo are able to cool resulting in 
disk galaxies which are too massive. 

Numerical simulations also show another shortcoming of the hierarchical clustering
hypothesis, which is to some extent related to the overcooling problem: the
angular momentum of simulated galaxies is too small. Compared to their
observed counterparts galaxies are thus too concentrated. This is shown in
figure \ref{ns97} which shows the specific angular momentum of dark
matter halos and of their central gaseous disks at $z=0$, as a function of
mass. The boxes indicate the loci corresponding to spiral and elliptical
galaxies, as compiled by Fall (1983). If, as suggested by Fall \& Efstathiou
(1980), the collapse of gas would proceed under conservation of angular
momentum, the baryonic component would have the same specific angular momentum $J/M$
as the dark matter, however, its corresponding mass would be a factor of 20
smaller (for $\Omega_{\rm bary} = 0.05$, $\Omega_0=1$). These disks would be
located only slightly below the box for spiral galaxies. However, figure
\ref{ns97} demonstrates clearly that the spins of gaseous disks are about an
order of magnitude lower than that. This is a direct consequence of the
formation process of the disks (Navarro, Frenk \& White 1995). Most of the disk mass is
assembled through mergers between systems whose own gas component had previously
collapsed to form centrally concentrated disks. During these mergers, and
because of the spatial segregation between gas and dark matter, the gas
component transfers most of their orbital angular momentum to the surrounding
halos (Frenk \etal, 1985; Barnes 1988; Quinn \& Zurek 1988).

As already mentioned, efficient feedback processes are required to
prevent the gas from excessive cooling. One proposal has been that energy input due
to a photoionizing UV background may prevent cooling in low mass halos at higher 
redshift (Efstathiou 1992). This hypothesis has been tested by numerical simulations but came to a 
negative result (Quinn, Katz \& Efstathiou 1996; Weinberg, Hernquist \& Katz
1997; Navarro \& Steinmetz 1997). These simulations assumed the presence of a
photoionizing background with an energy distribution 

\begin{equation}
J(\nu) = J_{-21}\times
10^{-21}\,\left(\frac{\nu}{\nu_{\rm H}}\right)^{-\alpha}\, .
\end{equation}

Though an UV background can delay or even prevent the formation of galaxies with
circular velocities of 30-50\,km/s and below, its influence on the properties of
galaxies with circular velocities exceeding 100\,km/s is almost
negligible. The amount of cool gas is only moderately reduced by about 10-30\%.
For extreme assumptions on the UV background it can be reduced 
by 50\%, insufficient to reconcile the
observed shape of the galaxy luminosity function.  Concerning the angular
momentum of gaseous disks, a UV background even exacerbates the angular momentum
problem as shown by the solid squares and starred symbols in figure \ref{ns97}.
This can be easily understood, since gas which falls in late and thus has low
densities is most strongly affected by the UV background. 
Such gas, however, also possesses the highest specific angular momentum.

To solve the overcooling/angular momentum problem thus energy feedback from
supernovae has been repeatedly advocated (\eg, Dekel \& Silk 1986). The stumbling
block for implementing star formation into galaxy formation simulations is
certainly the ill-understood physics of star formation and the interaction of
evolving stars with the ISM. The negative effect of photoheating on the angular
momentum of cold disk also points to an interesting complication which seem to
be generic to quite a variety of feedback mechanisms: In order to explain the
observed sizes of disk galaxies, the specific angular momentum of the disk must
not be much smaller than that of the host dark matter halo. However, it is also 
gas at large radii which (i) possesses the highest specific angular momentum and
which (ii) has the lowest density and for which cooling can be easily
suppressed. It is thus a non-trivial problem how the amount of cool gas can be
reduced by a large fraction without affecting its specific angular
momentum. This problem seems to be especially severe for low-$\Omega_0$ models,
where only a small fraction of baryons can be allowed to cool.
In the following sections, some results from simulations including star
formation and feedback are being discussed. These simulations incorporate
supernova feedback due only to thermal energy (see appendix). To some extent they can thus be
considered as minimum feedback models.

\section{Simulation Sample}

\begin{figure}[ht]
{\hskip 0.01 \hsize\epsfxsize=0.98\hsize\epsffile{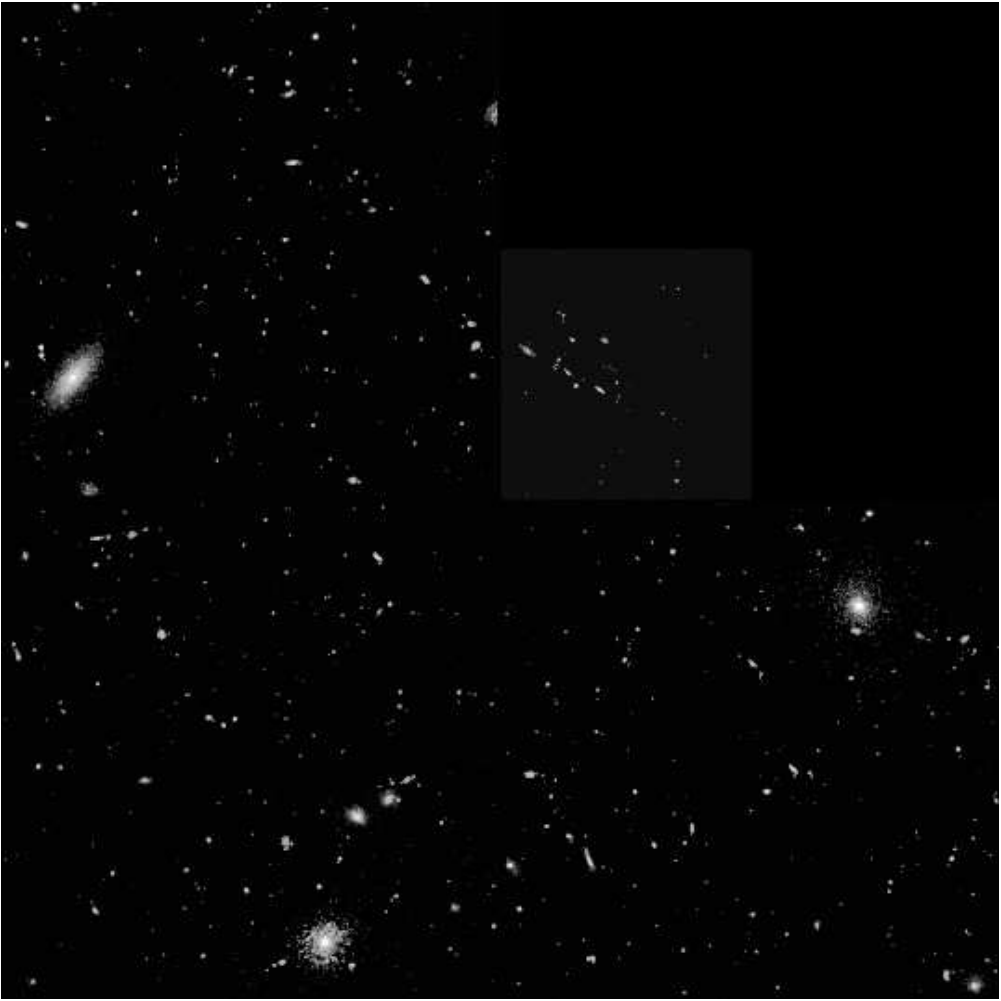}\hfill\break}
\caption{\label{fakehst}Computer-generated synthetic image of a WFPC2/PC field with
parameters chosen to match the exposures corresponding to the Hubble
Deep Field (HDF). All bands observed in the HDF have been simulated
and combined to produce this image. The cosmological model probed is
standard CDM.}
\end{figure}

The galaxy sample which is analyzed in the following sections consists of 21
galaxies with circular velocities between 50 and 250\,km/s. This sample has
been compiled from a set of 8 high resolution numerical simulations. The mass of
a gas particle is between $5\times 10^6$\,\Msol\ and $2\times 10^7$\,\Msol, the
gravitational softening is 1\,kpc. The star  formation efficiency has been calibrated
so that at $z=0$ a galaxy with a circular velocity of 200\,km/s has a star
formation rate of about 1\,\Msol/yr. Details of the numerical method and the star
formation and feedback scheme are presented in the appendix. Each galaxy
consists of $10^2-10^4$ star particles. Each of these particles
represents a population of a few million stars which have been formed in a
burst-like manner, \ie, a model galaxy can be considered as a superposition of
several thousand mini starbursts. Each star particle is labeled by an age
(time since creation) and a metallicity equal to that of its gas progenitor at
the time of formation.  The luminosity evolution of each of these bursts is then
followed by an evolutionary spectral synthesis model (Contardo, Steinmetz \&
Fritze-von Alvensleben 1998). The model galaxies can thus be ``observed'' in
arbitrary colors.  The power of this technique is illustrated in Figure
\ref{fakehst}, which shows a computer-generated rendition of a CDM-dominated
universe observed with the Hubble Space Telescope (HST) with exposures similar
to those used for the Hubble Deep Field (HDF). The image contains galaxies of
various intrinsic luminosities, placed at different redshifts and taken from the
simulation sample so as to reproduce approximately the apparent magnitude and redshift
distributions of galaxies in the HDF. The image includes realistic noise levels,
and has been processed with a point-spread function (PSF) similar to that of
HST. This image illustrates dramatically how simulations can be ``observed'' and
compared directly with high-resolution observations of distant galaxies.

\section{Tully-Fisher Relation}

\begin{figure}[ht]
{\hskip 0.01\hsize\epsfxsize=0.98\hsize\epsffile{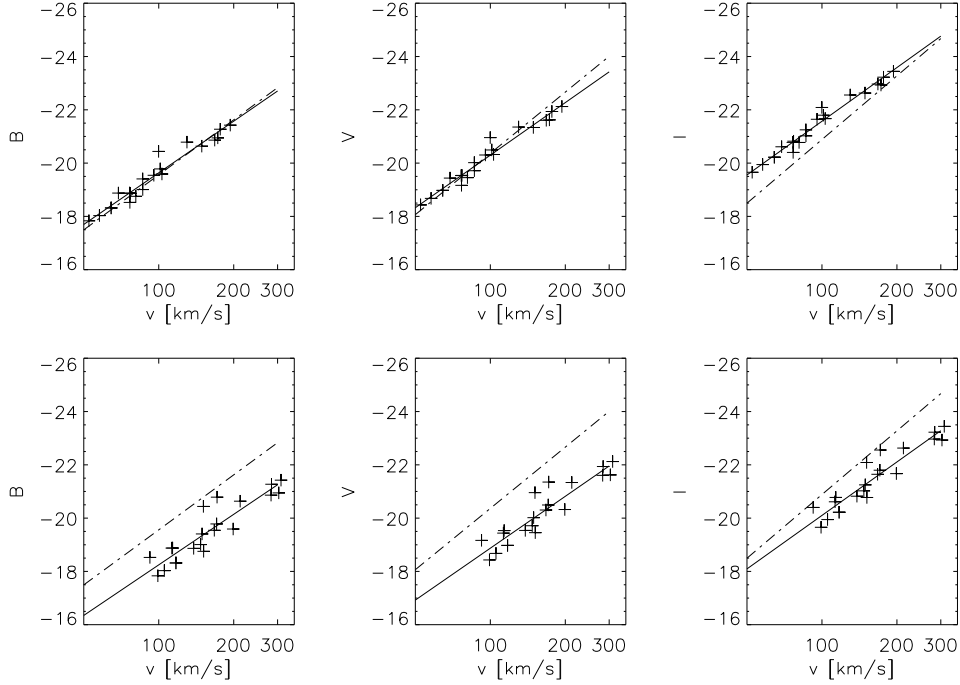}\hfill\break}
\caption{\label{tully}Tully-Fisher relation: B (left), V (middle) and I (right) luminosity of 
numerically simulated galaxies versus their circular velocity  at the virial
radius (top row) and their maximum rotation velocity (bottom row). The solid
lines correspond to a least square fit to the simulated data, the
dash-dotted lines correspond to the observed B,V and I Tully-Fisher relation (from
Pearce \& Tully, 1992).}
\end{figure}

The Tully-Fisher (TF) relation in several colors provides an excellent testbed for galaxy formation
simulations. The total baryonic mass of a dark-matter halo scales with its
circular velocity like $M\propto v^3$ (equation 2.2), and thus already 
provides a scaling similar to the observed
luminosity/velocity relation (see, however, Silk (1997) for a model in which the
TF relation arises mainly as a consequence of self-regulated star formation in galactic
disks). This basic relation is then modulated by the mass
distribution of the gas compared to the dark matter (\ie, the relation between
the actual rotation velocity of gas/stars and the virial velocity of the dark-matter 
halo), the efficiency of transforming gas into stars and the
mass-to-light ratios in different bands. All these modulations are likely to depend on the mass
(or, equivalently, circular velocity) of the halo and will affect
the normalization, scatter, and slope of the TF relation.  Hence, the kinematic of galaxies
(represented by their rotation velocities) is linked with
their current star formation rate (represented by their B-band luminosities), and their
star formation history (represented by their I or K-band
luminosities). Finally, these modulations have to preserve the small
scatter of about 0.3 mag in the observed TF relation. Further constraints 
can be derived by comparing the redshift evolution of the TF relation with
observations (\eg, Vogt \etal, 1997).

Figure \ref{tully} shows the B-, V- and I-band TF relation for the numerically
simulated galaxies. In order to demonstrate different contributions, the B, V and 
I luminosities are plotted against the circular velocity of the DM halo and against
the maximum velocity of the rotation curve. For comparison a best fit to the
simulated data (solid line) and the observed data (dashed line, from Pierce \&
Tully 1992) is
shown. The plot demonstrates that the simulations can qualitatively reproduce
many features of the observed data: first of all, a clear steepening of the TF
relation from B to I is visible. While the luminosity is proportional to
$L_B\propto v^{2.4}$, the I 
luminosity follows $L_I \propto v^{2.7}$. Also the {\it rms} scatter is remarkably small,
0.2 mag, if using the  virial velocity and 0.4
mag using the maximum rotation velocity. These data are consistent with
the observed values of  $\Delta M = 0.3$ mag (recall, dark matter halos obey 
$\Delta M = 0$ mag  and $M\propto v^3$.). Slope and scatter of the TF relations as well as
differences between different bands are a result of the simulations. 
The calibration of the star formation law
only enters by fixing the B luminosity of a galaxy within a halo of circular
velocity 200\,km/s !

However, a more quantitative comparison shows that the model fails in
detail. First of all, the slope of the TF relation is too flat, especially in the 
I band ($L_I \propto v^{2.7}$ versus observed $ L_I \propto v^{3.2}$) which
probably indicates that the supernovae feedback is not acting efficiently
enough. A more efficient feedback mechanism would deplete the total amount of
stars at low circular velocities and thus result in a lower I-band luminosities
and a steepening of the I-band TF relation. The systematically low luminosities
in the lower row of figure \ref{tully} can be interpreted in two ways: either
the luminosities are in fair agreement, but the galaxies are too concentrated
(\ie, the
maximum of the rotation curve is too high), or the velocities are in fair
agreement, but the luminosities are too low. The angular momentum problem
mentioned in the prologue supports the first interpretation, and also strengthens 
the conclusion that feedback has not worked efficiently enough. 
If, however, the actual rotation velocities of galaxy disks are more similar to
the virial velocities of its dark matter halo, the 
I band TF relation (figure \ref{tully}, upper right) indicates, that the total
amount of stars in a galaxy of given rotation velocity is too large. 

These results can be summarized as follows: the adopted feedback mechanism is
too weak to resolve the overcooling/angular momentum problem. The resulting
galaxies are too massive and too concentrated compared to their observed
counterparts. This effect is especially
strong at the low mass end and the slope of the I-band TF relation is too
shallow.  This discussion, however, also demonstrates how global scaling
relations like the TF relation can be used to calibrate models of (large-scale)
star formation and feedback in galaxy formation simulations.

\section{High Redshift Galaxies}

\begin{figure}[ht]
{\hskip0.01\hsize\epsfxsize=0.98\hsize\epsffile{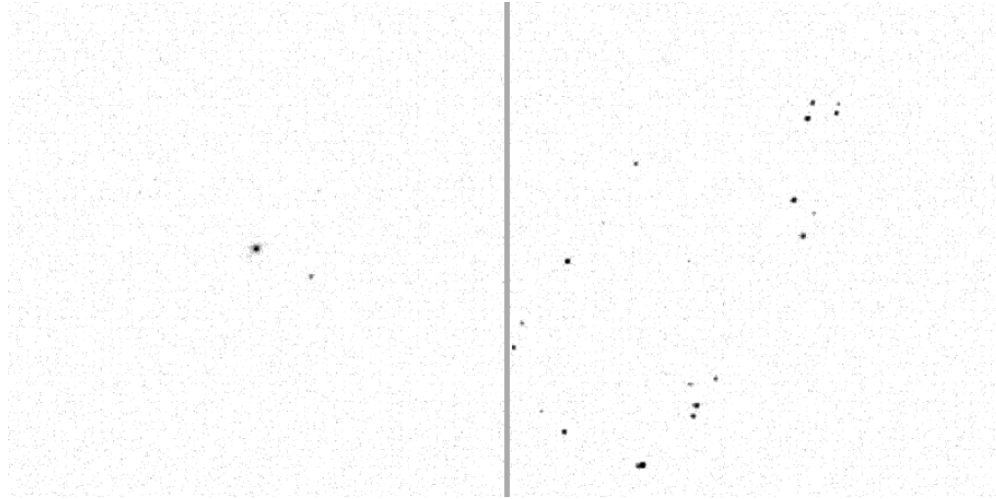}\hfill\break}
%
%\vspace{6.0cm}
\caption[]{\label{galred}Left: Artificial I-band image of a galaxy at redshift
z=0 as seen from a distance corresponding to $z=3$. Right: Progenitors of this
galaxies at redshift $z=3$. The group at the lower left edge will merge to one
galaxy outside the field covered by the left panel. The group near the center
merges and forms the two galaxies which can be seen in the left panel.
Resolution, Noise, PSF and efficiency are
taken to match that of the HST WFPC2 camera, exposure time: 123.6 ksec. 
Each frame has a sidelength of 2.8\,Mpc (comoving), corresponding to the area
covered by 4 WFPC2 chips.}
\end{figure}

Although the models still fail to  reproduce quantitatively the scaling relations of
present day galaxies, the qualitative agreement is probably good
enough to take a closer look at the redshift evolution of these galaxies,
especially if one concentrates on the high mass end where the influence of
feedback processes is weaker.

Figure \ref{galred} shows a $z=0$ galaxy and its progenitor at $z=3$, a group of
protogalactic clumps (PGCs, see also Haehnelt, Steinmetz \& Rauch 1996). The
baryonic masses of these clumps are only a few times $10^9$\,\Msol\ or even
less. Their mutual separation is about a few hundred kpc. All PGCs
share virtually the same redshift ($\Delta v\approx 400$\,km/s).  Similar to
the galaxy shown in figure \ref{galred}, most of the 21 simulated galaxies
give rise to a few detectable ($I<26$) progenitors at $z\approx 2-3$, a
behavior which nicely accounts for the increasing evidence for redshift
clustering at redshifts above two (Pascarelle \etal, 1996; Steidel \etal, 1997;
Elston \& Bechtold, 1998).

\begin{figure}[ht]
{\hskip 0.01 \hsize\epsfxsize=0.98\hsize\epsffile{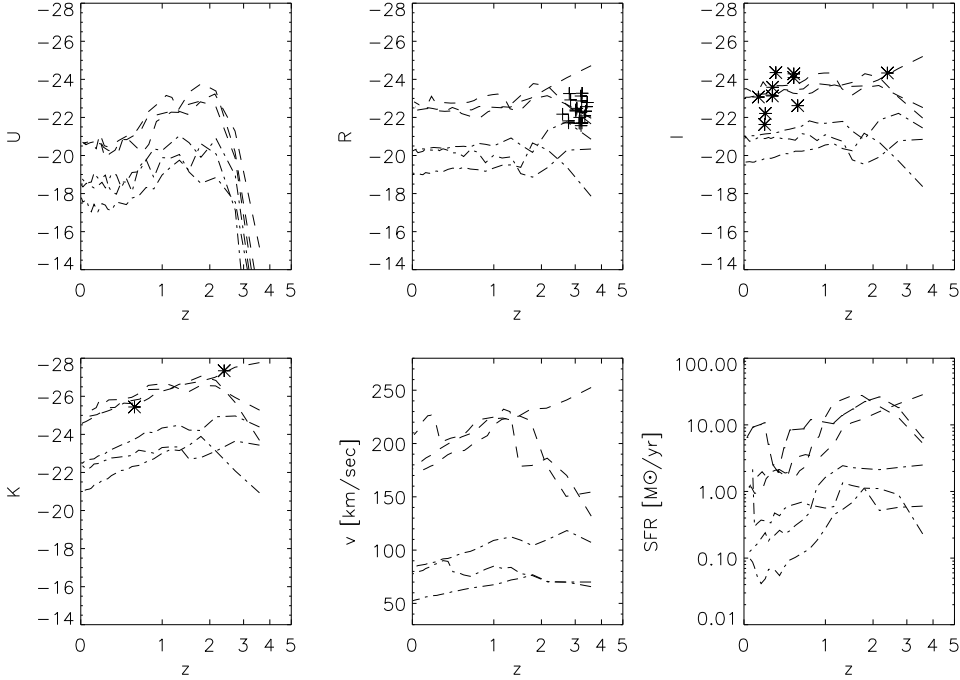}\hfill\break}
\caption{\label{prop} Upper row: redshift evolution of the absolute magnitude 
in U (left) R (middle) and I (right) of
three galaxies with $v_c(z=0) \approx 190\,$km/s (dashed) and three galaxies
with $v_c(z=0) \approx 80\,$km/s (dashed dotted). For comparison, ($+$) and
($*$) denoted  observational
data from Steidel \etal\ (1996) and Windhorst \etal\ (1994), respectively. Lower
row: redshift evolution in K (left), circular velocity as a function of redshift
(middle) and star formation rate as a function of redshift (right).}
\end{figure}

The redshift evolution of a subset of galaxies (3 galaxies with $v_c\approx
200\,$km/s and 3 galaxies with $v_c\approx 80\,$km/s) is shown in figure
\ref{prop}. By following the merging history of a galaxy, the
expression "progenitor" is, of course, no longer uniquely determined for redshifts higher
than that of last major merging event. Hence, at each branch in the merging tree, the
more massive clump is defined as the progenitor. Luminosity, velocity and star
formation rate shown in figure \ref{prop} always refers to only one bound clump.

The absolute U magnitude (figure \ref{prop}, upper left) brightens with redshift
about 1-2 magnitudes between redshift 0 and 2 which indicates a higher star
formation rate at higher $z$. This assumption is confirmed by the lower left
plot of figure \ref{prop} which shows the star formation rate of these
clumps. At $z=0$, the star formation rate is a very few 
\Msol/yr for 200\,km/s galaxies and an order of magnitude lower for 100\,km/s
galaxies. The star formation rate increases with redshift, however it
always stays below a few tens of \Msol/yr.
Note, however, that the simulations do not account for the effects of dust.

At redshift above 2.5 a strong drop in U is visible. This dropout is
mainly due to intervening absorption due to neutral hydrogen shortward of
$1217\,$\AA (restframe) and, to a smaller extent, absorption within the
atmospheres of young massive stars.  In R and I, the luminosity is
fairly constant. The plots also indicate, that the luminosities of the more
massive models are in good agreement with that of high redshift galaxies observed
by Windhorst \etal\ (1994) and Steidel \etal\ (1996).  Also the circular velocity
of these clumps is fairly constant. The circular velocity of low $v_c$ objects is even slightly
higher at larger redshifts. Also the circular velocity of the 200\,km/s
galaxies increases slightly with redshift for $z<1.5$, and it is still above
150\,km/s at $z=3$. So it should not be surprising that galaxies with velocities 
of a few hundred km/s exist at these redshifts (see
also Steinmetz 1997; Baugh \etal\ 1997).

In summary, the $z\la 3-4$ progenitors of present day galaxies have
luminosities and rotation velocities similar to those of their present day
counterpart. However, they are much more compact and much less massive.

\section{Absorption Systems}

\begin{figure}[ht]
{\hskip0.0\hsize\epsfxsize=0.48\hsize\epsffile{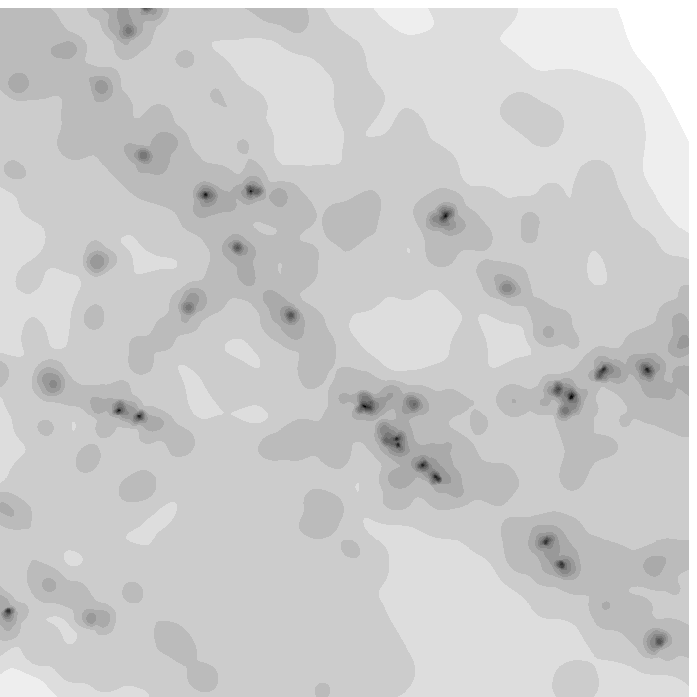}\hskip0.04\hsize\epsfxsize=0.48\hsize\epsffile{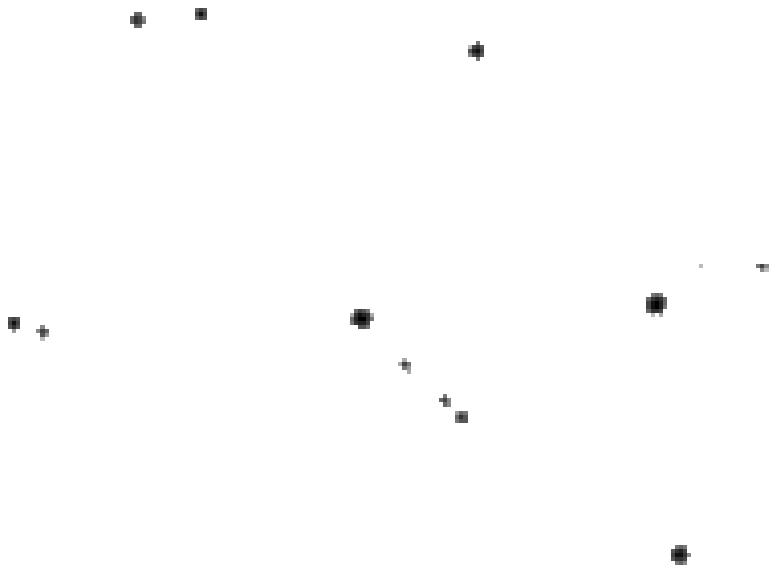}\break}
%
%\vspace{6.0cm}
\caption[]{\label{galcol}Right: \HI\ column density map of a galaxy forming
environment at at redshift z=3. Light gray correspond to column densities of
about  $\log N= 13.5$,
dark gray to $\log N \approx 15.5$ and black to  $\log N > 17.5$.
Left: the same simulation shown as an artificial I-band image  in an HDF-like
exposure. Each frame has a side length of 2.8\,Mpc (comoving).}
\end{figure}

One very successful application of gasdynamical simulations in cosmology has
been the study of the intergalactic medium (IGM), in particular the physical 
origin of  QSO absorption systems. Hydrodynamical simulations similar to those
presented here explain the basic properties of QSO absorbers
covering many orders of magnitude in \HI\ column density (see, \eg, Cen \etal, 1994;
Katz \etal, 1996; Zhang \etal, 1995; Rauch, Haehnelt \& Steinmetz 1997).
This is also shown in figure \ref{galcol} (left). While the lowest column
density systems ($\log N(\HI) \approx 12-14$, light gray) 
arises from gas in voids and sheets of the ``cosmic web'', systems of higher
column density are produced by filaments ($\log N \approx 14-17$, dark gray) or even gas
which has cooled and collapsed in virialized halos ($\log N > 17$, black).  

So far, numerical simulations have been  applied primarily to systems with lower 
column densities ($\log N \la 17$), corresponding to gas densities below
$10^{-2}\,$cm$^{-3}$.  At such low densities the important physical processes are 
relatively simple and well understood. Fluctuations are still only mildly
non-linear and the gas is essentially in photoionization  equilibrium with the 
UV background. Cooling times are long compared to dynamical time scales.

The right panel of figure \ref{galcol} shows the corresponding I-band image 
of the stellar component. The I band image includes
noise, PSF and exposure time similar to that of the Hubble Deep Field. The
artificial image shows about 8 detectable PGCs. Each of
these PGCs is situated close to a region of very high column density ($\log N >
17$). However, there is still a substantial number of Ly-limit and damped
\Lya systems, which do not host a stellar PGC.

There has been considerable debate about the physical structures giving rise to
damped \Lya absorption systems (DLAS) at high redshift. DLASs have often been
interpreted as large, high-redshift progenitors of present-day spirals which have
evolved little apart from forming stars (Wolfe 1988). More recently,
Prochaska \& Wolfe (1997) have argued that only models in which the
lines-of-sight (LOS) intersects rapidly rotating thick galactic disks can
explain both the large velocity spreads (up to 200\,km/s) and the
characteristic asymmetries of the observed low ionization species (\eg, SiII)
absorption profiles.  In particular, they find that if they embed their disk
model within a CDM structure formation scenario, the result is inconsistent with
the observed velocity widths. However, although Prochaska \& Wolfe investigated
quite a number of different geometrical and dynamical configurations, all models
have in common that the underlying mass distribution  is highly
symmetric and the models are in dynamical equilibrium.

\begin{figure}[ht]
{\hskip 0.01\hsize\epsfxsize=0.98\hsize\epsffile{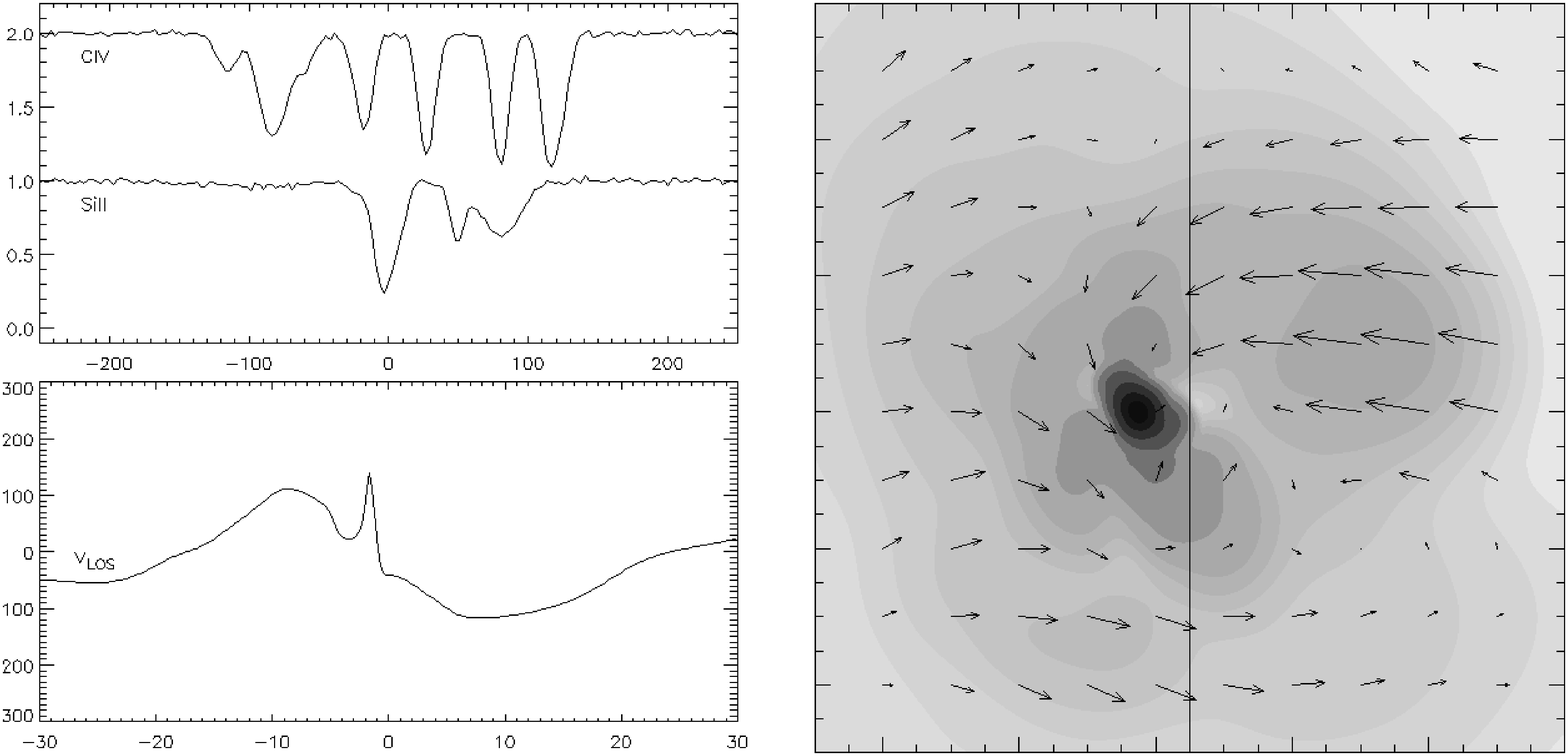}\hfill\break}
\caption{\label{damped}Right: Color map of the column density distribution in a
60\,kpc around a damped system. Black correspond to \HI\ densities $\log n (\HI)
> 1.5$), light grey to $\log n (\HI) \approx -3$). White arrows indicate the velocity field. The
white line correspond to the line-of-sight(LOS). In the lower left plot, the velocity
field along the LOS is shown. The upper left plot shows the absorption line in
CIV 1548 (top) and SiII 1808 (bottom). For readability, CIV has been displaced
by 0.5 in flux.}
\end{figure}

The importance of asymmetries and non-equilibrium effects has been demonstrated by
Haehnelt, Steinmetz \& Rauch (1998). Figure \ref{damped} shows a typical
configuration which gives rise to a high redshift DLAS with an asymmetric SiII
absorption profile.  The velocity width of about 120\,km/s is also quite similar
to typical observation.  However, no large disk has yet been developed and also
the circular velocity of the collapsed object is only 70\,km/s.  The physical
structure which underlies DLASs are turbulent gas flows and inhomogeneous
density structures related to the merging of two or more clumps, rather than
large rotating disks similar to the Milky Way. Rotational motions of the gas
play only a minor role for these absorption profiles.  A more detailed analysis
also demonstrates that the numerical models easily pass the statistical tests
proposed by Prochaska and Wolfe, \ie, hierarchical clustering, in particular the
CDM model, is consistent with the kinematics of high-$z$ DLASs. Semianalytical
models failed since they are based on the assumption of relaxed disk-like
structures rather than the complicated infall pattern seen in the simulations.

\section{Summary and Conclusion} 

Numerical simulations of the formation of galaxies through hierarchical
clustering have been presented. The simulation outcome has been compared
to a large variety of observations for galaxies at low and high redshifts. 
The main conclusions are the following:

\begin{itemize}

\item Hierarchical clustering has proven to be a very successful model for
structure formation in the universe. It also provides well defined initial
conditions from which the formation of galaxies can be studied. Some
observations like, \eg, the increasing evidence for redshift clustering are
natural predictions of the hierarchical clustering hypothesis.

\item Numerical simulations are an extremely powerful tool to study the
formation of galaxies. Only numerical simulation can take full account of the
dynamic of the formation process and the complicated interplay between different 
physical processes such as, \eg,
accretion and merging, star formation and feedback, photo heating and radiative 
cooling.

\item Overall, the qualitative picture seems to be fairly consistent, although
current models still fail to reproduce the properties of the observed galaxy
population in a quantitative manner. Future investigation will have to cope with 
the effects of (large-scale) star formation and feedback processes.
\end{itemize}

\begin{acknowledgments}
This article includes work from collaborations with
G.~Contardo, M.~Haehnelt, J.~Navarro and M.~Rauch. 
\end{acknowledgments}

\begin{appendix}
\section{Simulating Galaxy Formation}

\subsection{Numerical method}
The simulation presented in this article have been performed using GrapeSPH
(Steinmetz 1996), a particle method that combines the hardware  $N$-body
integrator GRAPE (=GRAvity PipE, Sugimoto \etal, 1990) with the
Smooth Particle Hydrodynamics (SPH) 
approach to numerical hydrodynamics (Lucy 1977; Gingold \& Monaghan 1977).  
GrapeSPH is fully Lagrangian, free from symmetry
restrictions, and highly adaptive in space and time through the use of
individual particle timesteps and smoothing lengths.  
The physical processes implemented in this code include self-gravity,
gas pressure, hydro-dynamical shocks, radiative cooling and heating by
a photoionizing background, and star formation.  

\subsection{Star formation}
Star formation is modeled by the creation of collisionless
``star'' particles in regions where the gas is locally Jeans unstable
and where the cooling timescale is shorter than the local dynamical
timescale. The local star formation rate per unit
volume is assumed to be directly
proportional to the gas density and inversely
proportional to a local dynamical timescale. A star formation efficiency of 3\%
has been used (for details see Katz 1992; Navarro \& White 1993; Steinmetz \& M\"uller
1994, 1995), a value consistent with that observed in the Milky Way.

The orbits of newly formed stars are subsequently followed in a
self-consistent fashion, assuming that they are only affected by
gravitational forces. Young star particles devolve energy and metal enriched
mass to their surrounding gas, an effect that mimics the energy and mass input
by supernovae and evolving stars into the ISM. The supernova energy is added to
the thermal energy of the surrounding gas. Input in kinetic energy (Navarro \&
White 1993) has not been assumed, \ie, this star formation model represents a
minimum feedback model.

\subsection{Initial conditions and simulation design}

The large dynamic range needed to resolve the internal structure of
galaxies and the full cosmological context of the galaxy formation
process is achieved using a two-step procedure. First, galaxy-sized
halos are extracted from large cosmological N-body
simulations. Second, these systems are resimulated in high-resolution
individual runs that use the same initial conditions and tidal fields
of the original simulations plus small scale perturbations introduced
to account for the increased Nyquist frequency of the second
run (Navarro, Frenk \& White 1995).

\end{appendix}


\begin{thebibliography}{} 
%
\bibitem[]{} Barnes, J. 1988, \apj, 331, 699.
%
\bibitem[]{} Baugh, C., Cole, S., Frenk, C.S., Lacey, C., 1998, \apj, submitted
(astro-ph/9703111)
%
\bibitem[]{} Cen, R., Miralda-Escud\'e, J., Ostriker, J.P., Rauch, M.,
1994, \apj,437, L9
%
\bibitem[]{} Contardo, G., Steinmetz, M., Fritze-von Alvensleben, U., 1998, in preparation.
%
\bibitem[]{} Cole, S.M., Arag\'on-Salamanca, A., Frenk, C.S., Navarro, J.F., Zepf, S.E.
~1994, \mnras, 271, 781.
%
\bibitem[]{} Dekel, A., \& Silk, J. 1986, \apj, 303, 39.
%
\bibitem[]{} Efstathiou, G.P. 1992, \mnras, 456, 43p.
%
\bibitem[]{} Elston, R., Bechtold, J., 1998 in preparation.
%
\bibitem[]{} Fall, S.M. 1983, in {\it Internal Kinematics and Dynamics of Galaxies}, Athanassoula E. (ed.), (Dordrecht: Reidel), p. 391.
%
\bibitem[]{} Fall, S.M. \& Efstathiou, G. 1980, \mnras, 193, 189.
%
\bibitem[]{} Frenk, C.S., White, S.D.M., Efstathiou, G.P., and Davis, M. 1985, \nat,
317, 595.
%
\bibitem[]{} Gingold, R.A., Monaghan, J.J., 1977 \mnras, 481, 375.
%
\bibitem[]{} Haehnelt, M., Steinmetz, M., Rauch, M., 1996, \apj, 465, L95.
%
\bibitem[]{} Haehnelt, M., Steinmetz, M., Rauch, M., 1998, \apj, in press (astro-ph/9706201).
%
\bibitem[]{} Katz, N., 1992, \apj, 391, 502.
%
\bibitem[]{} Katz, N., Weinberg, D. H., Hernquist, L., \& Miralda-Escud\`e J. 1996,\apj, 457, L57 
%
\bibitem[]{} Kauffmann, G., Nusser, A., Steinmetz, M., 1997, \mnras, 286, 795.
%
\bibitem[]{} Kauffmann, G., White, S.D.M., \& Guiderdoni, B. 1994, \mnras, 267, 981.
%
\bibitem[]{} Lucy, L., 1977, \aj, 82, 1013.
%
\bibitem[]{} Navarro, J.F., Frenk, C.S., \& White, S.D.M. 1995, \mnras, 275, 56.
%
\bibitem[]{} Navarro, J.F., Steinmetz, M., 1997, \apj, 471, 13.
%
\bibitem[]{} Navarro, J.F., White, S.D.M., 1993, \mnras, 265, 271.
%
\bibitem[]{} Pascarelle, S.M., Windhorst, R.A., Keel, W.C., Odewahn, S.C., \nat, 383, 45.
%
\bibitem[]{} Pierce, M.J., Tully, R.B., 1992, \apj, 387, 47.
%
\bibitem[]{} Prochaska, J. X., \& Wolfe, A. M. 1998, \apj, in press, (astro-ph/9704169)
%
\bibitem[]{} Quinn, P.J. \& Zurek, W.H. 1988, \apj, 331, 1.
%
\bibitem[]{} Quinn, T., Katz, N. \& Efstathiou, G. 1996, \mnras, 278, L49.
%
\bibitem[]{} Rauch, M., Haehnelt, M.G., Steinmetz, M., 1997, \apj 481, 601.
%
\bibitem[]{} Silk, J., 1997, \apj, 481, 703.
%
\bibitem[]{} Songaila, A.; Cowie, L. L., Hogan, C. J., Rugers, M., 1994, \nat, 368, 599.
%
\bibitem[]{} Steidel, C., Giavalisco M.,  Pettini, M., Dickinson, M., Adelberger, K., 1995, \apj, 462, L17
%
\bibitem[]{} Steidel, C., Adelberger, K., Dickinson, M., Giavalisco M., Pettini,
M., Kellogg, M.,1997, \apj, in press.
%
\bibitem[]{} Steinmetz, M., 1996, \mnras, 278, 1005.
%
\bibitem[]{} Steinmetz, M., 1997, {\sl Numerical Simulations of Galaxy Formation},
Proc.~ {\sl  Science with the VLT}, 
Garching, Germany, April 1 - 4 1996, Springer Verlag Berlin Heidelberg, 156.
%
\bibitem[]{} Steinmetz, M., M\"uller, E., 1994, \aaa, 281, L97.
%
\bibitem[]{} Steinmetz, M., M\"uller, E.,  1995, \mnras,  276, 549.
%
\bibitem[]{} Sugimoto, D., Chikada, Y., Makino, J., Ito, T., Ebisuzaki, T. \& Umemura, 
M. 1990, Nature, 345, 33.
%
\bibitem[]{} Tytler, D., Fan, X.-M., Burles, S., 1996, \nat, 381, 207.
%
\bibitem[]{} Vogt, N., \etal, 1997, \apj, 479, L121.
%
\bibitem[]{} Weinberg, D., Hernquist, L. \& Katz, N. 1997, \apj, 477, 8.
%
\bibitem[]{} Windhorst, R., \etal, 1994, \apj, 435, 577.
%
\bibitem[]{} White, S.D.M., 1994, Les Houches Lectures
%
\bibitem[]{} White, S.D.M., Rees, M.J. 1978, \mnras, 183, 341.
%
\bibitem[]{} White, S.D.M., Navarro, J.F., Evrard, A.E., Frenk, C.S., 1993,
\nat,366, 429.
%
\bibitem[]{} Wolfe, A. M. 1988, in {\it QSO Absorption Lines: Probing the Universe},
Proc. of the QSO Absorption Line Meeting, Baltimore, 1987, Cambridge 
University Press
%
\bibitem[]{} Zhang Y., Anninos P., Norman M.L., 1995, \apj, 453, L57 

\end{thebibliography}
\end{document}